\newcommand\apjcls{1}
\newcommand\aastexcls{2}
\newcommand\othercls{3}

\newcommand\papercls{\aastexcls}
\documentclass[tighten,times,twocolumn]{aastex62}  

\if\papercls \apjcls
\usepackage{apjfonts}
\else\if\papercls \othercls
\usepackage{epsfig}
\usepackage{margin}
\usepackage{times}
\fi\fi
\usepackage{ifthen}
\usepackage{natbib}
\usepackage{bm}
\usepackage{amssymb, amsmath}
\usepackage{appendix}
\usepackage{etoolbox}
\usepackage[T1]{fontenc}
\usepackage{paralist}
\usepackage{newtxtext,newtxmath}
\if\papercls \apjcls
\newcommand\aas{\ref@jnl{AAS Meeting Abstracts}}
\newcommand\dps{\ref@jnl{AAS/DPS Meeting Abstracts}}
\newcommand\maps{\ref@jnl{MAPS}}
\else\if\papercls \othercls
\usepackage{astjnlabbrev-jh}
\fi\fi

\if\papercls \aastexcls
\hypersetup{citecolor=blue, 
            linkcolor=blue, 
            menucolor=blue, 
            urlcolor=blue}  
\else
\usepackage[
bookmarks=true,           
bookmarksnumbered=true,   
colorlinks=true,          
citecolor=blue,           
linkcolor=blue,           
menucolor=blue,           
urlcolor=blue,            
linkbordercolor={0 0 1},  
pdfborder={0 0 1},
frenchlinks=true]{hyperref}
\fi

\if\papercls \othercls

\else

\fi

\providecommand{\adsurl}[1]{\href{#1}{ADS}}

\makeatletter
\patchcmd{\NAT@citex}
  {\@citea\NAT@hyper@{%
     \NAT@nmfmt{\NAT@nm}%
     \hyper@natlinkbreak{\NAT@aysep\NAT@spacechar}{\@citeb\@extra@b@citeb}%
     \NAT@date}}
  {\@citea\NAT@nmfmt{\NAT@nm}%
   \NAT@aysep\NAT@spacechar\NAT@hyper@{\NAT@date}}{}{}

\patchcmd{\NAT@citex}
  {\@citea\NAT@hyper@{%
     \NAT@nmfmt{\NAT@nm}%
     \hyper@natlinkbreak{\NAT@spacechar\NAT@@open\if*#1*\else#1\NAT@spacechar\fi}%
       {\@citeb\@extra@b@citeb}%
     \NAT@date}}
  {\@citea\NAT@nmfmt{\NAT@nm}%
   \NAT@spacechar\NAT@@open\if*#1*\else#1\NAT@spacechar\fi\NAT@hyper@{\NAT@date}}
  {}{}
\makeatother

\makeatletter
\DeclareRobustCommand{\lowcase}[1]{\@lowcase#1\@nil}
\def\@lowcase#1\@nil{\if\relax#1\relax\else\MakeLowercase{#1}\fi}
\makeatother

\DeclareSymbolFont{UPM}{U}{eur}{m}{n}
\DeclareMathSymbol{\umu}{0}{UPM}{"16}
\let\oldumu=\umu
\renewcommand\umu{\ifmmode\oldumu\else\math{\oldumu}\fi}

\if\papercls \othercls

\else

\fi

\let\oldsim=\sim
\renewcommand\sim{\ifmmode\oldsim\else\math{\oldsim}\fi}
\let\oldpm=\pm
\renewcommand\pm{\ifmmode\oldpm\else\math{\oldpm}\fi}
\newcommand\by{\ifmmode\times\else\math{\times}\fi}

\newbox{\wdbox}
\renewcommand\c{\setbox\wdbox=\hbox{,}\hspace{\wd\wdbox}}
\renewcommand\i{\setbox\wdbox=\hbox{i}\hspace{\wd\wdbox}}

\newcount\timect
\newcount\hourct
\newcount\minct
\newcommand\now{\timect=\time \divide\timect by 60
         \hourct=\timect Cltiply\hourct by 60
         \minct=\time \advance\minct by -\hourct
         \number\timect:\ifnum \minct < 10 0\fi\number\minct}

\catcode`@=11

\newcommand\comment[1]{}

\newcommand\commenton{\catcode`\%=14}

\renewcommand\math[1]{$#1$}
\newcommand\mathshifton{\catcode`\$=3}

\let\atab=&
\newcommand\atabon{\catcode`\&=4}

\let\oldmsp=\sp
\let\oldmsb=\sb
\def\sp#1{\ifmmode
           \oldmsp{#1}%
         \else\strut\raise.85ex\hbox{\scriptsize #1}\fi}
\def\sb#1{\ifmmode
           \oldmsb{#1}%
         \else\strut\raise-.54ex\hbox{\scriptsize #1}\fi}
\newbox\@sp
\newbox\@sb
\def\sbp#1#2{\ifmmode%
           \oldmsb{#1}\oldmsp{#2}%
         \else
           \setbox\@sb=\hbox{\sb{#1}}%
           \setbox\@sp=\hbox{\sp{#2}}%
           \rlap{\copy\@sb}\copy\@sp
           \ifdim \wd\@sb >\wd\@sp
             \hskip -\wd\@sp \hskip \wd\@sb
           \fi
        \fi}
\def\msp#1{\ifmmode
           \oldmsp{#1}
         \else \math{\oldmsp{#1}}\fi}
\def\msb#1{\ifmmode
           \oldmsb{#1}
         \else \math{\oldmsb{#1}}\fi}

\def\supon{\catcode`\^=7}

\def\subon{\catcode`\_=8}

\def\supsubon{\supon \subon}

\newcommand\actcharon{\catcode`\~=13}

\newcommand\paramon{\catcode`\#=6}

\comment{And now to turn us totally on and off...}

\newcommand\reservedcharson{ \commenton  \mathshifton  \atabon  \supsubon 
                             \actcharon  \paramon}

\catcode`@=12
\reservedcharson

\if\papercls \apjcls

\else

\fi

\newcommand\chisq{\ifmmode{\chi\sp{2}}\else\math{\chi\sp{2}}\fi}
\newcommand\redchisq{\ifmmode{ \chi\sp{2}\sb{\rm red}}
                    \else\math{\chi\sp{2}\sb{\rm red}}\fi}
\newcommand\Teq{\ifmmode{T\sb{\rm eq}}\else$T$\sb{eq}\fi}
\newcommand\mjup{\ifmmode{M\sb{\rm Jup}}\else$M$\sb{Jup}\fi}
\newcommand\rjup{\ifmmode{R\sb{\rm Jup}}\else$R$\sb{Jup}\fi}
\newcommand\msun{\ifmmode{M\sb{\odot}}\else$M\sb{\odot}$\fi}
\newcommand\rsun{\ifmmode{R\sb{\odot}}\else$R\sb{\odot}$\fi}
\newcommand\mearth{\ifmmode{M\sb{\oplus}}\else$M\sb{\oplus}$\fi}
\newcommand\rearth{\ifmmode{R\sb{\oplus}}\else$R\sb{\oplus}$\fi}

\newcommand{\Ro}{\ensuremath{\mathrm{R}_{\rho}}}

\newcommand\Pran{\ensuremath{\mathrm{Pr}}}

\renewcommand{\bar}[1]{\overline{#1}}
\newcommand{\grad}{\bm{\nabla}}

\shorttitle{Rotation reduces compositional mixing in Jupiter and other gas giants}
\shortauthors{Fuentes {\em et al.}}

\begin{document}

\title{Rotation reduces convective mixing in Jupiter and other gas giants}


\author{J. R. Fuentes}
\affiliation{\rm Department of Applied Mathematics, University of Colorado Boulder, Boulder, CO 80309-0526, USA}

\author{Evan H. Anders}
\affiliation{\rm Center for Interdisciplinary Exploration and Research in Astrophysics, Northwestern University, Evanston, Illinois 60201, USA
}

\author{Andrew Cumming}
\affiliation{\rm Department of Physics and Trottier Space Institute, McGill University, Montreal, QC H3A 2T8, Canada}

\author{Bradley W. Hindman}
\affiliation{\rm Department of Applied Mathematics, University of Colorado Boulder, Boulder, CO 80309-0526, USA}
\affiliation{\rm JILA, University of Colorado Boulder, Boulder, CO 80309-0440, USA}


\begin{abstract}

Recent measurements of Jupiter's gravitational moments by the Juno spacecraft and seismology of Saturn's rings suggest that the primordial composition gradients in the deep interior of these planets have persisted since their formation. One possible explanation is the presence of a double-diffusive staircase below the planet's outer convection zone, which inhibits mixing across the deeper layers. However, hydrodynamic simulations have shown that these staircases are not long-lasting and can be disrupted by overshooting convection. In this letter, we suggests that planetary rotation could be another factor for the longevity of primordial composition gradients. Using rotational mixing-length theory and 3D hydrodynamic simulations, we demonstrate that rotation significantly reduces both the convective velocity and the mixing of primordial composition gradients. In particular, for Jovian conditions at $t\sim10^{8}~\mathrm{yrs}$ after formation, rotation reduces the convective velocity by a factor of 6, and in turn, the kinetic energy flux available for mixing gets reduced by a factor of $6^3\sim 200$. This leads to an entrainment timescale that is more than two orders of magnitude longer than without rotation. We encourage future hydrodynamic models of Jupiter and other gas giants to include rapid rotation, because the decrease in the mixing efficiency could explain why Jupiter and Saturn are not fully mixed.

\end{abstract}


\section{Introduction}

The history of the solar system is written in the interiors of the giant planets. 
In particular, since Jupiter accounts for 75\% of the Solar System's planetary mass, understanding its interior (or more precisely, determining the distribution and total mass of heavy elements within the planet) can provide important clues on the first stages of the formation of the solar system \citep[see the recent review by][]{2023RemS...15..681M}.
Traditionally, Jupiter has been considered a fully convective planet, with a well-defined interface separating a dense core made of heavy elements and the hydrogen-helium envelope \citep{1996Icar..124...62P}. 
However, new data from the Juno mission indicates that Jupiter does not have a well-defined central core of heavy elements, nor is it homogeneously mixed \citep[e.g.,][]{2017SSRv..213....5B,2017Sci...356..821B,2017GeoRL..44.4649W,2019ApJ...872..100D,2023A&A...672A..33H}. 
Interestingly, ring seismology suggests that Saturn is likewise not homogeneously mixed \citep{2021NatAs...5.1103M}. 
These new observational constraints on the internal structure of gas giants require updates to formation and evolution models \citep{2022Icar..37814937H}.  
Composition gradients decrease the efficiency of the heat transport throughout the planet’s interior and thus affect its cooling, and mass-radius relationship at a given age \citep[e.g.,][]{2007ApJ...661L..81C,2012A&A...540A..20L,2013NatGe...6..347L}.

Primordial composition gradients are expected to naturally result from planet formation \citep[e.g.,][]{2017ApJ...840L...4H,2020A&A...638A.121M,2022PSJ.....3...74S}  or collisional events during the planet's evolution \citep{2019Natur.572..355L}, but the persistence of composition gradients against strong convective-mixing over evolutionary timescales is poorly understood. 
One mechanism for maintaining a composition gradient appears in one-dimensional evolution models, where convective staircases (i.e., multiple convective layers separated by thin, stably-stratified, diffusive interfaces) form underneath the outer convective envelope, preventing mixing deeper in the interior \citep[e.g.,][]{2015ApJ...803...32V,2018A&A...610L..14V,2022PSJ.....3...74S}.  However, 3D simulations of layered convection in incompressible flows have shown that convective staircases fully mix on short timescales \citep[e.g.,][]{2013ApJ...768..157W,2018AnRFM..50..275G}. Recently, \cite{2022PhRvF...7l4501F} explored the formation of a convective staircase beneath a growing convection zone,  and found that compositional mixing across the interface between the convection zone and the stable region prevents the formation of the staircase. Similar results were found by \citet{2022ApJ...928L..10A} in the context of stellar convection, concluding that compositional gradients are ineffective barriers against convective mixing over evolutionary timescales.

Rotation is another factor that could prevent convective mixing in Jupiter and other gas giants. 
Experiments and numerical simulations of thermal convection have shown that rotation reduces the convective velocities when the flow has small Rossby number, $\mathrm{Ro}$, the ratio of the rotation period to the convective turnover time \citep[e.g.,][]{1979GApFD..12..139S, 1991JFM...228..513F,2014ApJ...791...13B,2018FrEaS...6..189D,Aurnou2020,2020ApJ...898..120H}. 
Further, numerical experiments wherein rapidly-rotating convection zones grow via entrainment  have shown that the vertical transport of buoyancy and kinetic energy across the convective layer, and thus the entrainment rate, are reduced by rotation \citep{1996DyAtO..24..237J}. 
Convective flows in Jupiter and other gas giants can be significantly slow compared to rotation \citep[$\mathrm{Ro}\lesssim 10^{-6}$, e.g.][]{2004jpsm.book...35G}, so the effects of rotation on the erosion of primordial composition gradients are likely to be significant. Indeed, \cite{2012A&A...540A..20L} suggested that rotation would hamper convection and impede the mixing of heavy elements.

In this Letter, we examine how rotation modifies the rate of mixing of a stable fluid layer by a neighboring convection zone. In Section~\ref{sec:MLT}, we use mixing-length theory to argue that rotation reduces the convective velocity and the kinetic energy flux available for mixing. Then, in Section~\ref{sec:entrainment} we use the scalings from rotating mixing length theory to show that rotation decreases the rate of entrainment and mixing at the boundary between the stable and unstable layers. In Section~\ref{sec:experiment_and_results}, we use 3D numerical simulations to support the predictions from the previous sections. Finally, in Section~\ref{sec:disc} we discuss the implications of our results for Jupiter and other gas giants.

\section{Mixing-length theory for rotating convection}

In this section, we summarize the mixing-length scalings for the convective velocity and kinetic energy flux in non-rotating and rotating flows. 

\label{sec:MLT}

In the absence of rotation, convective motions are nearly isotropic with a  characteristic length scale, $\ell$, that is roughly the depth of the convection zone (e.g.,~in a laboratory experiment) or a pressure scale height (e.g.,~in stellar and planetary interiors). 
Further, the velocity scale of the convective flows is approximately the overturning buoyant ``free-fall'' velocity, $U_{\mathrm{NR}} = \sqrt{g\alpha \delta T_\mathrm{NR} \ell}$, where the subscript $\mathrm{NR}$ means \emph{no rotation}, $g$ is the acceleration due to gravity, and $\alpha = -(\partial \ln \rho/\partial T)|_C$ is the coefficient of thermal expansion at constant composition $C$, and $\delta T_{\mathrm{NR}}$ is the characteristic temperature perturbation of convective parcels. 
The heat flux carried by the convective motions is $F_H \sim \rho c_P \delta T_\mathrm{NR} U_{\mathrm{NR}}$, where $\rho$ is the mass density, and $c_P$ is the specific heat capacity at constant pressure. Then, the convective velocity can be estimated as
\begin{equation}
U_{\mathrm{NR}}\sim \left(\dfrac{\alpha g F_H \ell}{\rho c_P}\right)^{1/3}\,. \label{eq:no_rot}
\end{equation}
Typical values for Jupiter's deep interior at $t\sim 10^{8}~\mathrm{yrs}$ after formation, when the convective luminosity is high enough so that a large fraction of the planet is expected to be mixed, $\ell \sim H_P\sim 10^9~\mathrm{cm}$ (a pressure scale height), $c_P/\alpha g \sim 10^{10}~\mathrm{cm}$, $\rho\sim 1~\mathrm{g~cm^{-3}}$, and $F_H\sim 6\times 10^{5}~\mathrm{cgs}$ \citep[e.g.,][]{2018MNRAS.477.4817C,2022Icar..37814937H}, give $U_{\mathrm{NR}}\sim 40~\mathrm{cm\,s^{-1}}$ (and accordingly, the convective turnover time is $\ell/U_{\mathrm{NR}} \sim 0.8~\mathrm{yr}$).

In the limit of rapid rotation, convection is quasi-geostrophic, and convective flows are set by a balance between the Coriolis, inertial, and buoyancy (Archimedean) forces \citep[called CIA balance, e.g.,][]{1979GApFD..12..139S}. In this limit, a new length scale, $\ell_{\perp}$, associated with motions perpendicular to the rotation vector, $\bm{\Omega}$, emerges \citep[e.g.,][]{2014ApJ...791...13B,Aurnou2020,2021PNAS..11822518V}. 
From CIA balance, one obtains
\begin{equation}
\dfrac{2\Omega U_{\mathrm{R}}}{\ell} \sim \dfrac{U^2_{\mathrm{R}}}{\ell^2_{\perp}} \sim \dfrac{\alpha g \delta T_\mathrm{R}}{\ell_{\perp}}\,, \label{eqn:cia}
\end{equation}
where the subscript R means \emph{rotating} and $\Omega$ is the angular frequency of the rotation vector. Using the definition of the heat flux $F_H \sim \rho c_P \delta T_\mathrm{R} U_{\mathrm{R}}$, and eliminating $\delta T_{\mathrm{R}}$ and $\ell_{\perp}$ from Equations~\eqref{eqn:cia}, it follows that
\begin{align}
U_{\mathrm{R}}
\sim \left(\dfrac{g\alpha F_H}{\rho c_P}\right)^{2/5}\left(\dfrac{\ell}{2\Omega}\right)^{1/5}\, 
\sim U_{\rm NR}\left(\frac{U_{\rm NR}}{2 \Omega \ell}\right)^{1/5} ,\label{eq:u_rot}
\end{align}
Plugging in the previous values for Jupiter, together with $2\pi/\Omega\sim 10$ hours, we obtain $U_{\mathrm{R}} \sim 6~\mathrm{cm\,s^{-1}}$. 
Rotation therefore reduces the convective velocity by a factor of $\sim 6$; then, the convective turnover time increases by a factor of $\sim 6.6$, giving $\ell/U_{\mathrm{R}} \sim 5~\mathrm{yrs}$. 

Although the effect of rotation on the convective velocity could be considered minor in the sense that the resulting turnover time is still many orders of magnitude smaller than the age of the planet, the important quantity for entrainment is the kinetic energy flux, which is reduced by a much larger factor. This can be seen from the ratio between the kinetic energy flux $F_{\rm K} \sim (1/2)\rho U_{\mathrm{conv}}^3$ associated with the convective motions and the thermal heat flux carried by convection. Without rotation, Equation~\eqref{eq:no_rot} gives $F_{\mathrm{K,NR}}/F_H = (1/2)\rho U_{\mathrm{NR}}^3/F_H \sim (\alpha g \ell/c_P)$.  In the limit of rapid rotation, Equation~\eqref{eq:u_rot} gives
\begin{equation}
\frac{F_{\mathrm{K,R}}}{F_H} \sim \left(\frac{\alpha g \ell}{c_P}\right)\left(\frac{U_{\mathrm{R}}}{2\Omega \ell}\right)^{1/2} \sim  \mathrm{Ro}^{1/2}\frac{F_{\rm K,NR}}{F_H}\,,  \label{eq:flux_ratio}
\end{equation}
where $\mathrm{Ro} = U_{\mathrm{R}}/2\Omega \ell \ll 1$ is the Rossby number of the flow based on the rotationally-constrained convective velocity. Then, \emph{for a given heat flux}, rotation reduces the kinetic energy flux by a factor of $\mathrm{Ro}^{1/2} \ll 1$. Using the values above for Jupiter, we estimate $\mathrm{Ro}\sim 10^{-5}$, and consequently, the kinetic energy flux available to mix the fluid is reduced by a factor of $\sim 4\times 10^{-3}$. This reduction is important because the energy source for the upward transport and mixing of heavier material is the kinetic energy of the convective motions. We investigate this in more detail in Section~\ref{sec:entrainment}.

\section{Entrainment in rapidly rotating flows} \label{sec:entrainment}

Mixing of material (entrainment) across convective boundaries (e.g., a core-envelope interface) has been investigated with great detail in laboratory experiments and numerical simulations \citep[e.g.,][]{1968JFM....33..183T,1987JFM...182..525F,molemaker97,2020PhRvF...5l4501F}. 
In those studies, a convection zone advances into a stable region, and the returning flows from overshooting convection entrain fluid with the chemical composition of the stable layer. 
This material rapidly mixes, so the convection zone grows over time \citep[see][for a review of boundary mixing processes]{anders_pedersen_2023}. 
By reducing the convective velocities, rotation reduces the entrainment (mixing) efficiency, and consequently, decreases the growth rate of the convective layer. 

Consider a plane-parallel fluid layer of depth $H$, stably-stratified with composition decreasing linearly with height, $C_0(z) = |dC_0/dz|(H-z)$. Note that $C$ could represent the spatial variation in mass fraction of heavy elements, or the mean molecular weight of the fluid, as would be induced by salt in water or heavy elements in hydrogen.
Then, a heat flux $F_0$ escapes through the top of the stable layer, such that convection is driven from above and mixes the composition gradient from top to bottom. For simplicity, we adopt the Boussinesq approximation \citep{1960ApJ...131..442S}, wherein the fluid density is constant everywhere except for buoyant perturbations, which is a decent approximation in the vicinity of a convective boundary.

We now estimate the rate of entrainment $dh/dt$, where $h(t)$ is the depth of the convection zone at time $t$.
Lifting and mixing a quantity $\Delta C$ of heavy fluid requires an energy expenditure $\Delta E = (1/6)\rho_0\beta \Delta C g h^2$  \citep[see, e.g., ][]{2020PhRvF...5l4501F}, where mass conservation dictates that $\Delta C = 0.5|dC_0/dz|h$. 
Here, $\rho_0$ is the density of the layer, and $\beta = (\partial \ln \rho/\partial C)_T$ is the coefficient of compositional expansion (both quantities assumed constant).
Differentiating $\Delta E$ with respect to time, we write the rate of change of potential energy due to mixing as $d(\Delta E)/dt = (1/4)\rho_0 \beta |dC_0/dz| g h^2\dot{h}$.
We adopt the entrainment hypothesis, which states that the rate of change of potential energy across the layer is set by the kinetic energy flux of the convective motions $F_{\mathrm{KE}} = (1/2)\rho_0 U^3_{\mathrm{conv}}$ \citep{1975JFM....71..385L}. It follows that
\begin{equation}
\dfrac{1}{4}\rho_0 \beta g \bigg| \dfrac{dC_0}{dz}\bigg|h^2\dfrac{dh}{dt} \approx \dfrac{1}{2}\rho_0 U^3_{\mathrm{conv}}\, .\label{eq:entrainment}
\end{equation}
Equation~\eqref{eq:entrainment} assumes that all the kinetic energy is used to do work and lift material from below, and that thermal effects do not significantly affect the rate of change in the potential energy across the layer (a good approximation since $\alpha \Delta T \ll \beta \Delta C$).

We see that given $U_{\mathrm{conv}}$, Equation~\eqref{eq:entrainment} can be integrated to obtain $h(t)$. Without rotation, we use  $\rho=\rho_0$, $F_H=F_0$, and $\ell=h$ and assume $U_{\mathrm{conv}} \approx U_{\mathrm{NR}} \propto \ell^{1/3}$ (Equation~\ref{eq:no_rot}), to obtain (within a factor of order unity due to the assumptions above)\footnote{A full derivation including the mixing efficiency, interfacial heat transport, and thermal effects on the density can be found in \citet{2020PhRvF...5l4501F}.}
\begin{equation}
h_{\mathrm{NR}}(t) \approx 2\left(\dfrac{\alpha g F_0}{\rho_0 c_P N^3}\right)^{1/2}\left(Nt\right)^{1/2}\,,\label{eq:h_no_rot}
\end{equation}
where $N = \sqrt{g\beta|dC_0/dz|}$ is the buoyancy frequency. From Equation~\eqref{eq:h_no_rot}, the rate at which the convection zone expands inward depends on how quickly it cools down (which sets the convective velocity magnitude) and on the magnitude of the initial composition gradient.

On the other hand, in the limit of rapid rotation we should use $U_{\mathrm{conv}} \approx U_{\mathrm{R}} \propto \ell^{1/5}$ (Equation~\ref{eq:u_rot}), and integration of Equation~\eqref{eq:entrainment} gives
\begin{equation}
h_{\mathrm{R}}(t) \approx \left(\frac{24}{5}\right)^{5/12}\left(\dfrac{\alpha g F_0}{\rho_0 c_P N^3}\right)^{1/2}\left(\dfrac{N}{2\Omega}\right)^{1/4} \left(Nt\right)^{5/12}\, ,\label{eq:h_rot}
\end{equation}
or
\begin{equation}
    h_{\rm R}(t) \approx \left[0.96 \left(\dfrac{N}{2\Omega}\right)^{1/4} (N\,t)^{-1/12}\right] h_{\rm NR}(t)\,.
\end{equation}
We can see that rotation slows down the propagation of the convection zone, it reduces the power-law index from 0.5 to approximately 0.41, but the most significant effect comes from the prefactor ($N\,t)^{-1/12}$, which can be very small over long timescales.

It is instructive to compare the mixing timescales in both cases. From Equations~\eqref{eq:h_no_rot} and~\eqref{eq:h_rot}, the time required for convection to mix a stable layer of vertical size $L$ (i.e.,~$h(t)=L)$ is 
\begin{equation}
\tau_{\mathrm{NR}}\approx N^{-1}\left(\dfrac{\rho_0 c_P N^3 L^2}{4\alpha g F_0}\right)\,, \label{eq:t_mix_nr}
\end{equation}
in the non-rotating case, and
\begin{align}
\tau_{\mathrm{R}} &\approx \dfrac{5}{24} N^{-1}\left(\dfrac{\rho_0 c_P N^3 L^2}{\alpha g F_0}\right)^{6/5}\left(\dfrac{2\Omega}{N}\right)^{3/5}
\approx \dfrac{5}{6}\mathrm{Ro}^{-1/2}~\tau_{\mathrm{NR}}\, ,\label{eq:t_mix}
\end{align}
in the rotating case. Therefore, the mixing timescale increases approximately by a factor of $(5/6)\mathrm{Ro}^{-1/2}$, a commensurate change to the decrease in the kinetic energy flux---compare Equations~ \eqref{eq:flux_ratio} and \eqref{eq:t_mix}.

In a system like Jupiter where $\mathrm{Ro} \sim 10^{-5}$ at $t\sim 10^{8}~\mathrm{yrs}$ after formation, this corresponds to a mixing timescale that is more than two orders of magnitude longer. This difference is substantial, e.g., for $L\sim 10^9~\mathrm{cm}$, $N^2\sim g/H_P\sim 10^{-6}~\mathrm{s^{-2}}$, Equations~\eqref{eq:t_mix_nr} and \eqref{eq:t_mix} give $\tau_{\mathrm{NR}}\sim 10^{8}~\mathrm{yrs}$ and $\tau_{\mathrm{R}}\sim 10^{10}~\mathrm{yrs}$, i.e., for the rotating case the mixing timescale would be more than the age of the Solar System.

\section{Numerical experiments and results} \label{sec:experiment_and_results}

\begin{figure*}
    \centering
    \includegraphics[width=\textwidth]{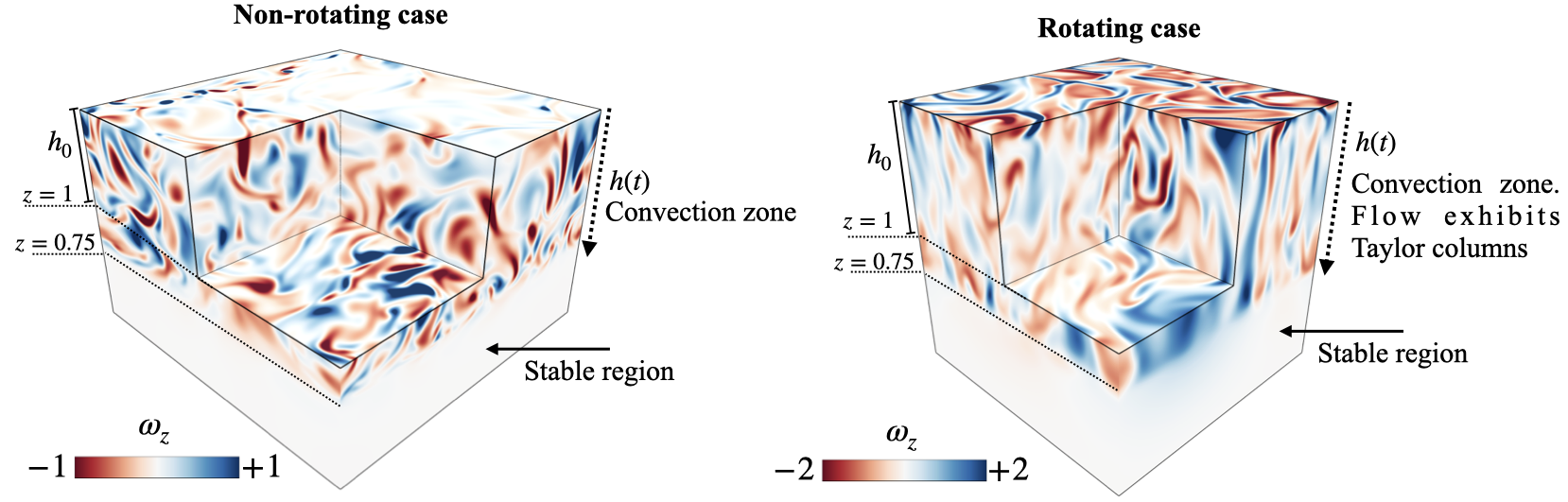}\\
    \caption{Volume renderings of the $z$ component of the vorticity $\bm{\omega} = \nabla \times \bm{u}$ for the non-rotating case (left) and rotating case (right). For both cases, the snapshots are shown at times when the bottom of the convection zone has expanded beyond the initial location (from $z = 1$ to 0.75). The Rossby number in the rotating case is $\mathrm{Ro_{sim}} \approx 0.065$, computed using the averaged flow velocity in the convection zone, and thickness of the convection zone.} \label{fig:dynamics}
\end{figure*}

\begin{figure*}
    \centering
    \includegraphics[width=\textwidth]{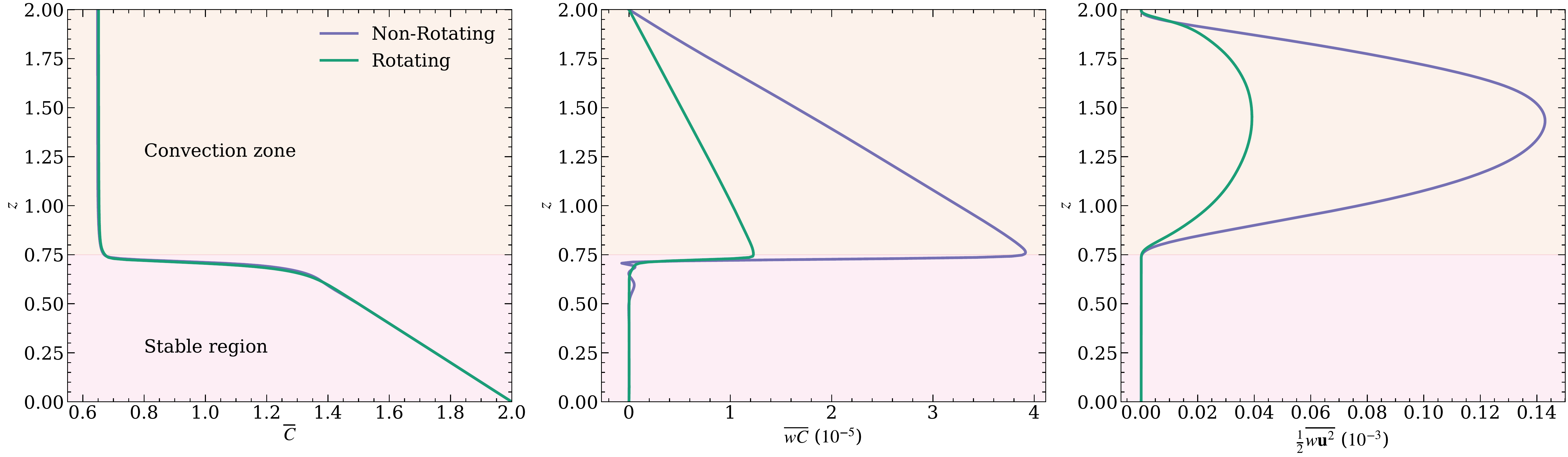}\\
    \caption{Horizontally-averaged profiles of composition $\overline{C}$, convective composition flux  $\overline{wC}$, and the kinetic energy flux in the vertical direction  $(1/2)\overline{w\bm{u}^2}$ (left, middle, and right panel, respectively). The overline denotes that variables are averaged over the horizontal directions. Note that $\bm{u}^2 = u^2 + v^2 + w^3$, where $u$, $v$, and $w$ are the  $x$, $y$, and $z$ components of the velocity. The results are shown at at a times where the convection boundary is located at $z=0.75$ (i.e., $h=1.25$) for both runs. This time corresponds to $t=3750$ for the non rotating case, and $t=11500$ for the rotating case. The background colors distinguish between the convection zone and the stable region.} \label{fig:profiles}
\end{figure*}

In the following, we use two hydrodynamic simulations (with and without rotation) to test the theory above. 
We model 3D thermal convection in a plane-parallel binary mixture of fluid (heavy and light elements). 
The domain initially consists of two layers: a convection zone of uniform composition, on top of a region stabilized by a composition gradient. 
The top of the convection zone is constantly cooled so that convection propagates downwards. We focus on how rotation affects the mixing of heavy material at the lower convective boundary.

In our simulations, as in our derivation above, we assume that the length scale of the fluid motions is much smaller than a density scale height, so that the effects of curvature can be ignored and the Boussinesq approximation is valid \citep{1960ApJ...131..442S}. 
Although the density in giant planets can vary by many orders of magnitude, this approximation fully captures nonlinear mixing near the convective boundary, which is the focus of our study. 
In the rotating case, we consider uniform rotation throughout the fluid, and assume that gravity and rotation are aligned and point in the z direction, $\bm{\Omega} = \Omega \hat{\bm{z}}$, and $\bm{g} = -g\hat{\bm{z}}$, i.e., the simulation is at a polar latitude. The density of the fluid depends on both composition and temperature fluctuations. 
In what follows, all results are presented in dimensionless form. 
For details on the numerical setup, nondimensionalization, and code, we refer the reader to Appendices~\ref{app:numerics_eqs} and ~\ref{app:simulation_details}.

Figure \ref{fig:dynamics} shows 3D snapshots of the $z$ component of the vorticity for both the non-rotating and rotating simulations. Initially, in both cases, the convection zone spans the upper half of the domain, $z \in [1,2]$. 
After significant evolution ($t\sim 3750$ for the non-rotating case, and $t\sim 11500$ for the rotating case), the bottom of the convection zone has advanced to $z = 0.75$, mixing the primordial composition gradient. 
Note that in the rotating case, it takes approximately 3 times longer for the zone to increase its vertical extent by $\Delta h = 0.25$. We show later that this difference agrees with Equation~\eqref{eq:t_mix}. Rotation visibly alters the flow morphologies. 
In the non-rotating case, the characteristic flow length scale is similar in all directions, while in the rotating case, the horizontal length scale of the flow is much smaller than the vertical scale \citep[as expected from the Taylor-Proudman theorem, e.g.,][]{1916RSPSA..92..408P,1917RSPSA..93...99T}. 

Figure~\ref{fig:profiles} shows horizontally-averaged profiles of the composition (left), vertical convective composition flux (middle), and vertical kinetic energy flux (right); fluxes are depicted at the same instant in time as displayed in Figure~\ref{fig:dynamics}. Rotation does not affect the composition profile, which is homogeneous in the convection zone with a transition to the primordial linear distribution in the stable region. This is expected because mass conservation requires that the amount of material transported and mixed in the convection zone depends only on the thickness of the convective layer, $\Delta C = 0.5|dC_0/dz|h$. 
The initial composition in the convection zone is $C = 0.5$, and with $|dC_0/dz| = 1$ and $\Delta h = 0.25$ giving $\Delta C = 0.125$, the expected evolved convection zone composition is $C = 0.625$, which is what we observe (see left panel). 
In the rotating case, both the convective composition flux and the kinetic energy flux are smaller by a factor of $\approx 3$ when compared with the values of the non-rotating case (see middle and right panels), supporting both our hypotheses that rotation decreases the vertical transport of dense material and that this reduction is related to the kinetic energy flux. Although we only focus on the compositional transport and kinetic energy flux, it is worth mentioning that the thermal structure of the fluid across the box is similar to that of the composition. The only difference is that in the non-rotating case, the convection zone tends to be isothermalized by strong turbulence, while in the rotating case, a small temperature gradient tends to be sustained in the convection zone \citep[this has been reported in previous studies of thermal convection in rotating flows, see, e.g.,][]{1996JFM...322..243J,2020PhRvF...5k3501C}.

\begin{figure}
    \centering
    \includegraphics[width=8cm]{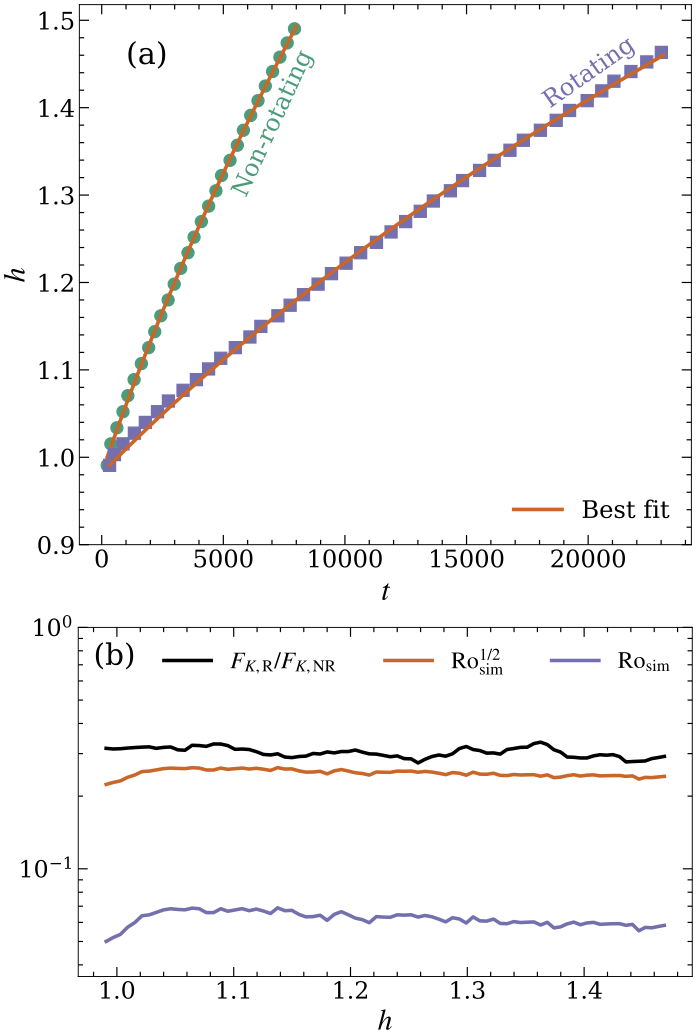}
    \caption{Panel (a): Thickness of the convection zone $h$ as a function of time $t$. The results are shown for both the rotating and non-rotating cases. The solid lines correspond to the best fits of the form $h(t) = [h_0^{1/n} + K(t-t_0)]^{n}$ for each case. We obtain $n\approx 0.49$ for the non-rotating case (compared to the analytic prediction $n = 0.5$) and $n \approx 0.40$ for the rotating case (compared to the analytic prediction $n = 5/12 \approx 0.41$). See text for more details. Panel (b): $\mathrm{Ro_{sim}}$, $\mathrm{Ro_{sim}}^{1/2}$, and $F_{K,\mathrm{R}}/F_{K,\mathrm{NR}}$, as a function of the thickness of the convection zone, $h$. Note that $\mathrm{Ro_{sim}}^{1/2}$ has similar values to $F_{K,\mathrm{R}}/F_{K,\mathrm{NR}}$, giving support to Equation~\eqref{eq:flux_ratio}.}  \label{fig:h_t}
\end{figure}

We expect from the entrainment rates calculated in Section~\ref{sec:entrainment} that the convection zone will grow more rapidly in the nonrotating case than the rotating case. 
We quantify this difference by measuring the size of the convection zone $h$ as a function of time. We define the bottom of the convection zone as the location where the convective composition flux is maximized (see Figure~\ref{fig:profiles}, middle), so that $h$ is the distance between that location and the top of the box. 
Figure~\ref{fig:h_t}~(a) shows $h$ vs.~time for the two cases; as expected, $h$ in the non-rotating case grows more rapidly than in the rotating case. We fit a power-law function of the form $h(t) = [h_0^{1/n} + K(t-t_0)]^{n}$, where $h_0$ is the size of the convection zone at initial time $t_0$, and $K$ is a constant. In both simulations, we demand that $h_0 = 1$, since initially the upper half of the box is mixed. The time $t_0$ is formally the time when the convection zone starts to grow.  This only occurs after the convective turbulence is fully established. The initial state in our simulations is one of quiescence. In the non-rotating case, the convection continues to form and equilibrate up to $t_0 = 225$ and only then does entrainment of the stable fluid begin. The convection in the rotating case takes longer to be established, $t_0 = 325$.
We find $K \approx 1.61\times 10^{-4}\, (6.62\times 10^{-5})$ and $n \approx 0.49\,(0.4)$ in the non-rotating (rotating) case. These values  are close to what is predicted by the simplified model in Section~\ref{sec:entrainment}, where the predicted powerlaw index is $n = 1/2 = 0.5$ for the non-rotating case, and $n = 5/12\approx 0.41$ for the rotating case. In dimensionless form, the predicted values of the prefactor are $K = 4(R_\rho/\mathrm{Pe_{ff}}) \sim 10^{-4}$ for the non-rotating case, and $K = (24/5) R_\rho( \mathrm{Ro_{ff}}/\mathrm{Pe_{ff}}^2)^{3/5} \sim 2\times 10^{-5}$ for the rotating case, where the input dimensionless parameters $R_\rho$, $\mathrm{Pe_{ff}}$, and $\mathrm{Ro_{ff}}$, are the density ratio, and the Peclet and Rossby numbers based on the free-fall velocity, respectively. The reader is refered to Appendix~\ref{app:numerics_eqs} for more details on the nondimensionalization.
 
Finally, we measure the Rossby number of the flow, $\mathrm{Ro_{sim}} = \lVert u \rVert/2\Omega h$, where $\lVert u \rVert$ and $h$ are the magnitude of the flow velocity averaged within the convection zone and the thickness of the convection zone, respectively. We find $\mathrm{Ro_{sim}}\approx 0.065$, which decreases slightly as the convection zone evolves and deepens (see Figure~\ref{fig:h_t}b). 
When compared with the ratio between the kinetic energy flux in the rotating and non-rotating case (as predicted by mixing-length theory, see Equation~\ref{eq:flux_ratio}), we find $\mathrm{Ro_{sim}}^{1/2}/(F_{K,\mathrm{R}}/F_{K,\mathrm{NR}}) \approx 0.87$. Finally, 
our theory (Equation~\ref{eq:t_mix}) predicts that mixing a depth $\Delta h$ should take longer in the rotating case by a factor of $(5/6)\mathrm{Ro}^{-1/2}$.  In our simulations this difference is approximately a factor of 3, close to $(5/6)\mathrm{Ro_{sim}}^{-1/2}\approx 3.3$.

\section{Summary and Discussion} \label{sec:disc}

In this work, we studied the effect of rotation on the compositional transport across the boundary between a convection zone and a stable region (e.g., a core-envelope interface in a gas giant). Mixing length theory predicts  that rotation reduces the convective velocity (compare Equations~\ref{eq:no_rot} and \ref{eq:u_rot}). For Jovian conditions at $t\sim 10^{8}~\mathrm{yrs}$ since formation, this corresponds to a velocity reduction of approximately a factor of 6. Consequently, the kinetic energy flux available for the entrainment and mixing of heavy material gets reduced by a factor of $(U_{\mathrm{NR}}/U_{\mathrm{R}})^3 \sim \mathrm{Ro}^{-1/2} \sim 6^3\sim 200$ (see Equation~\ref{eq:flux_ratio}). Using the prescriptions for the convective velocity from mixing-length theory, we derived a simple analytical model for the rate of entrainment and the propagation of a convection zone. The thickness of the outer convection zone follows $h(t)\propto t^{1/2}$ in the non-rotating case (Equation~\ref{eq:h_no_rot}), and $h(t)\propto \Omega^{1/4} t^{5/12}$ in the rotating case (Equation~\ref{eq:h_rot}), supporting the hypothesis that rotation decreases compositional mixing. 
From this model, the timescale to mix a stable layer of vertical size $L$ is larger by a factor of $(5/6)\mathrm{Ro}^{-1/2}$, where $\mathrm{Ro}$ is the Rossby number, when rotation is taken into account in the convective velocity (Equation~\ref{eq:t_mix}).
We ran two numerical experiments (including and not including rotation) of convective entrainment and found a good agreement between the analytic predictions and the simulations (e.g., see Figures~\ref{fig:profiles}--\ref{fig:h_t}).

The agreement between our numerical simulations and the mixing-length theory predictions is encouraging. The mixing time can increase by two orders of magnitude for Jupiter-like values at $t\sim 10^8$ years after formation. 
This could significantly reduce the rate of core erosion, preventing strong mixing without the need to invoke a double-diffusive staircase. For example, \cite{2017ApJ...849...24M} derived that the rate at which core mass is transported into the envelope is given by  $\dot{M}_{\mathrm{core}} = \gamma^{-1} \alpha L_{\mathrm{core}}/c_P$, where $L_{\mathrm{core}}$ is the luminosity at the core-envelope interface, and $\gamma^{-1}$ is ratio of the buoyancy flux associated with heavy-element transport to the buoyancy flux associated with heat transport. 
Using 3D simulations of non-rotating incompressible flows, \citet{2017ApJ...849...24M} constrained $\gamma^{-1} \gtrsim 0.5$ for Jupiter conditions. 
In the entrainment model investigated in this work, the rate at which mass is transported into the outer convection zone is $\dot{M}_{\mathrm{core}}\propto h dh/dt$. Then, it follows that rotation reduces the erosion rate by a factor of $\dot{M}_{\mathrm{core,R}}/\dot{M}_{\mathrm{core,NR}} = \gamma^{-1}_{\mathrm{R}}/\gamma^{-1}_{\mathrm{NR}} \approx \left(N/2\Omega\right)^{1/2}\left(24Nt/5\right)^{-1/6}$ (where we have used Equations~\ref{eq:h_no_rot} and \ref{eq:h_rot}). Using values for Jupiter at $t\sim 10^{8}$ years after formation, we estimate $\gamma^{-1}_{\mathrm{R}}/\gamma^{-1}_{\mathrm{NR}}\approx 0.01$. Although we have derived this ratio with a rather simple model, the fraction is significantly smaller than unity and illustrates the importance of rotation for convective boundary mixing. 




Although our numerical experiments are done in a Cartesian box and assume that gravity and rotation are parallel to each other (as expected at the poles of the planet), 
the reduction in the convective velocity does not depend on the exact geometry or shape of the composition profile. However, numerical simulations with misaligned gravity and rotation have shown that thermal mixing is not uniform across latitude \citep[e.g.,][]{2018FrEaS...6..189D,2020MNRAS.493.5233C}. A logical next step would be to investigate this problem using spherical geometry and a primordial composition gradient that is more consistent with formation models \citep[e.g.,][]{2022PSJ.....3...74S}.

Our simulations do not include density stratification, as we adopt the Bousinessq approximation which formally limits the vertical scale to be much less than a density scale height. 
Including density stratification and compressibility will change the mixing rate by a combination of two effects. 
First, the potential energy required to mix a region should decrease; while composition is uniformly mixed in the convection zone in both cases, the compressible case mixes the density to the adiabatic gradient, whereas Boussinesq convection mixes the density to a constant value, which requires a greater energy expenditure. 
Second, stratification increases the asymmetry between upflows and downflows and in turn increases the net kinetic energy flux, so it should matter whether a stable entraining region is at the top or bottom of a convection zone \citep[see e.g., Fig.~4][]{2007ApJ...667..448M}. 
At this time we cannot speculate on the net result of those effects. To our knowledge, there are no simulations of convective entrainment accounting for both composition gradients and rotation in a compressible flow. We will investigate this problem in a future paper. 

We encourage future work on hydrodynamic models of gas giants, as improved mixing prescriptions at convective boundaries provide a promising avenue for better aligning 1D evolution models with observational inferences of gas giant interiors from NASA's missions Juno and Cassini.

\begin{acknowledgements}
We thank Nick Featherstone and Keith Julien for useful conversations on rotating convection. A.C. acknowledges support by NSERC Discovery Grant RGPIN-2023-03620. J.R.F. and B.W.H. are supported by NASA through grants 80NSSC18K1125, 80NSSC19K0267 and 80NSSC20K0193. E.H.A. is supported by a CIERA Postdoctoral Fellowship. A.C. is a member of the Centre de Recherche en Astrophysique du Québec (CRAQ) and the Trottier Institute for Research on Exoplanets (iREx). Computations were conducted with support from the NASA High End Computing (HEC) Program through the NASA Advanced Supercomputing (NAS) Division at Ames Research Center on Pleiades with allocation GID s2276.
\end{acknowledgements}

\bibliography{references}{}
\bibliographystyle{aasjournal}

\appendix

\section{Fluid equations \& Initial Conditions} \label{app:numerics_eqs}

We study convective entrainment in rapidly rotating flows with composition and temperature gradients. To do so, we adopt the Boussinesq approximation of the fluid equations and evolve both temperature $T$ and composition $C$.  We normalize the 
fluid equations using as units of length and time the initial size of the convection zone $h_0$ and the free fall time across it $t_{\mathrm{ff}} = h_0/u_{\mathrm{ff}}$ (where $u_{\mathrm{ff}} = \sqrt{\alpha g h_0 \Delta T_F}$, $g$ is the acceleration due to gravity, and $\alpha \equiv -(\partial \ln \rho / \partial T)|_{C}$ is the coefficient of thermal expansion). Further, for the characteristic temperature scale we utilize $\Delta T_F = F_0 h_0/\rho_0\kappa_T c_P$, with $\rho_0$ being the background density, $F_0$ the cooling energy flux at the surface of the convection zone, $\kappa_T$ the thermal diffusivity, and $c_P$ the specific heat at constant pressure. 
The units of temperature and composition are respectively $[T] = \Delta T_F$, and $[C] = \Delta C_0$ (i.e., the initial composition change across the stable region). The corresponding normalizations for the kinetic, heat and composition fluxes are $[F_K] = (1/2)\rho_0 u_{\mathrm{ff}}^3$, $[F_H] = \rho c_P u_{\mathrm{ff}}\Delta T_F$, and $[F_X] = \rho_0 u_{\mathrm{ff}}\Delta C_0$, respectively. 

The nondimensional equations are
\begin{align}
    &\grad\cdot\bm{u} = 0 
        \label{eqn:incompressible} \\
    &\partial_t \bm{u} + \bm{u}\cdot\grad\bm{u} + \grad \varpi + \mathrm{Ro_{ff}}^{-1} \hat{\bm{z}} \times \bm{u} = \left(T - \frac{C}{\Ro}\right) \hat{\bm{z}} + \frac{\Pran}{\rm{Pe_{ff}}}\grad^2 \bm{u}
        \label{eqn:momentum}, \\
    &\partial_t T + \bm{u}\cdot \grad T = \frac{1}{\rm{Pe_{ff}}}\grad^2 T
        \label{eqn:temperature},\\
    &\partial_t C + \bm{u}\cdot\grad C = \frac{\tau_0}{\rm{Pe_{ff}}}\grad^2\bar{C} + \frac{\tau}{\rm{Pe_{ff}}}\grad^2 C',
        \label{eqn:composition}
\end{align}
where $\bm{u} = (u,v,w)$ is the velocity field, and $\varpi$ is the reduced pressure. Overbars denote horizontal averages and primes denote fluctuations around that average, such that e.g., $C = \bar{C} + C'$. 
Note that we reduce diffusion on $\bar{C}$ by choosing $\tau_0 < \tau$ to ensure its evolution is due to advection.

In the equations above, there are 5 dimensionless numbers that characterize the evolution of the flow. These are the freefall Rossby number, Peclet number, density ratio, Prandtl number, and inverse Lewis number, which are defined respectively as
\begin{align}
\begin{split}
    &\mathrm{Ro_{ff}} = \frac{u_{\rm{ff}}}{2 \Omega h_0} ,\hspace{0.1cm}
    \mathrm{Pe_{ff}} = \frac{u_{\rm{ff}} h_0}{\kappa_T},\hspace{0.1cm}
    R_\rho = \frac{\alpha\Delta T_F}{\beta\Delta C_0},\hspace{0.1cm}
    \Pran = \frac{\nu}{\kappa_T},\hspace{0.1cm}
    \tau = \frac{\kappa_C}{\kappa_T},
\end{split}
\end{align}
where $\Omega$ is the rotation rate, $\alpha \equiv (\partial \ln \rho / \partial T)|_{C}$ and $\beta \equiv (\partial \ln \rho / \partial C)|_{T}$ are respectively the coefficient of thermal and compositional expansion/contraction, $\nu$ is the kinematic viscosity, and $\kappa_T$ and $\kappa_C$ are the thermal and compositional diffusivity, respectively.
Note that $R_\rho$ is the ratio between the destabilizing and stabilizing effect of the thermal and compositional buoyancy.

The initial structure of the fluid is composed of two layers, defined across $z \in [0, 2]$, where the temperature and composition  are respectively
\begin{align}
    &T(z) = 2-z\, ,\qquad C_0(z) = 
    \begin{cases}
        0.5        & z > 1 \\
        2-z       & z < 1 \\
    \end{cases}.
    \label{eqn:initial_C}
\end{align}
Step functions are not well represented in pseudospectral codes, so we use a smoothed Heaviside function centered at $z = z_0$,
\begin{equation}
H(z; z_0, d_w) = \frac{1}{2}\left(1 + \mathrm{erf}\left[\frac{z - z_0}{d_w}\right]\right).
\label{eqn:heaviside}
\end{equation}
where erf is the error function; to construct $C_0(z)$ in Equation.~\ref{eqn:initial_C}, we set
\begin{equation}
    C_0(z) = (2 - z) [1 - H(z; 1, 0.05)] + \frac{1}{2} H(z; 1, 0.05). \label{eq:C}
\end{equation}
We fix the flux of temperature and composition at the top of the domain and we fix their values at the bottom of the domain, so the boundary conditions for temperature and composition are $\partial_z T(z=2) = -1$, $T(z=0) = 2$, 
$\partial_z C(z=2) = 0$, and 
$C(z=0) = 2$. 
The boundary conditions for the velocity are stress free at the upper boundary ($\hat{z}\cdot\bm{u} = \hat{x}\cdot\partial_z\bm{u} = \hat{y}\cdot\partial_z\bm{u} = 0$ at $z = 2$) and no-slip at the lower boundary in the stable fluid ($\bm{u} = \bm{0}$ at $z = 0$).

The simulations in this work use $\mathrm{Pe_{ff}} = 4.4 \times 10^3$, $R_{\rho}^{-1} = 10$, $\mathrm{Pr} = \tau = 0.5$,  and $\tau_0 = 10^{-3}$. We set the Coriolis force to be identically zero in the non-rotating case ($\mathrm{Ro_{ff}} = \infty$) , while in the rotating case $\mathrm{Ro_{ff}} = 0.81$. Under Jovian conditions, $\mathrm{Ro_{ff}} = \mathcal{O}(10^{-5})$, $\mathrm{Pe_{ff}} = \mathcal{O}(10^{10})$ $\tau \approx \mathrm{Pr} = \mathcal{O}(10^{-2})$, and $R^{-1}_\rho = \mathcal{O}(10)$ \citep[see e.g.,][]{2004jpsm.book...35G,2012ApJS..202....5F}, but we make clear that the density ratio is very uncertain given our ignorance on the compositional and thermal stratification in the deep interiors of the gas giants.)
Our simulations are as turbulent as possible and are qualitatively in the same regime as gas giants ($\mathrm{Ro_{ff}} < 1$, $\mathrm{Pe_{ff}} \gg 1$, $\mathrm{Pr} < 1$, $\tau < 1$, $R_\rho^{-1} > 1$). It is worth mentioning that although the input Rossby number $\mathrm{Ro_{ff}}$ is close to unity, the Rossby number computed from the simulated flows is much smaller ($\approx 0.065$). It would be possible to decrease the Rossby number, but the time that would be required for the convection zone to mix material would be significantly longer and beyond our computational resources.

\section{Simulation Details}
\label{app:simulation_details}
We time-evolve equations \ref{eqn:incompressible}-\ref{eqn:composition} using the Dedalus pseudospectral solver \citep{2020PhRvR...2b3068B} version 3\footnote{We use the code in the \texttt{master} branch of the \href{https://github.com/DedalusProject/dedalus}{\texttt{https://github.com/DedalusProject/dedalus}} repository, and use the commit with short-sha 29f3a59.} using timestepper SBDF2 \citep{wang_ruuth_2008} and CFL safety factor 0.2.
All variables are represented using a Chebyshev series with 512 terms for $z \in [0, L_z]$, with $L_z=2$, and Fourier series in the periodic $x$ and $y$ directions. In the non-rotating case, $x \in [0, L_x]$, $y \in [0, L_y]$, with $L_x=L_y = 3$, and each direction has 192 terms. In the rotating case, $L_x=L_y = 2$, and each direction has 128 terms. To avoid aliasing errors, we use the 3/2-dealiasing rule in all directions.
To start our simulations, we add random noise temperature perturbations sampled from a normal distribution with a magnitude of $10^{-5}$ to the initial temperature field.
The Python scripts used to run the simulations in this work are available online as a Git repository\footnote{\href{https://github.com/evanhanders/rotation_reduces_entrainment}{\texttt{https://github.com/evanhanders/rotation\_reduces\_entrainment}}, v1.0.}, and backed up in a Zenodo repository \citep{code_zenodo}.

\end{document}